\documentclass{IEEEtran}
\usepackage[hyphens]{url}
\usepackage{float, graphicx}
\usepackage{tabularx}

\title{Algorithm for Cross-shard Cross-EE Atomic User-level \eth Transfer in Ethereum}

\author{\IEEEauthorblockN{Raghavendra Ramesh}
\IEEEauthorblockA{\textit{Consensys Software R \& D} \\
Brisbane, Australia \\
raghavendra.ramesh@consensys.net}}

\newcommand{\eth}[0]{ETH~}
\newcommand{\tocredit}[0]{{\bf ToCredit}~}

\begin{document}

\maketitle

\begin{abstract}
Sharding is a way to address scalability problem in blockchain technologies. Ethereum, a prominent blockchain technology, has included sharding in its roadmap to increase its throughput. The plan is also to include multiple execution environments. 

We address the problem of atomic cross shard value transfer in the presence of multiple execution environments. We leverage on the proposed Ethereum architecture, more specificially on Beacon chain and crosslinks, and propose a solution on top of the netted-balance approach that was proposed for EE-level atomic \eth transfers. We split a cross-shard transfer into two transactions: a debit and a credit. First, the debit transaction is processed at the source shard. The corresponding credit transaction is processed at the destination shard in a subsequent block. We use {\em netted} shard states as channels to communicate pending credits and pending reverts. We discuss various scenarios of debit failures and credit failures, and show our approach ensures atomicity even in the presence of a Byzantine Block proposer. 

The benefits of our approach are that we do not use any locks nor impose any constraints on the Block Proposer to select specific transactions. However we inherit the limitation of an expensive operation from the netted-balance approach of querying partial states from all other shards. We also show a bound on the size of such inter-shard state reads.
\end{abstract}


\section{Introduction}

Ethereum~\cite{eth}, a widely adopted Blockchain platform, is pressed with the demand to increase its throughput. To address this scalability challenge, Ethereum community is moving towards a sharded~\cite{sharding} setup, where the global state of the system is distributed among different shards. There is a central coordinating chain called {\em Beacon chain} that serves as a coordinator of all the shards. The proposal is that the blocks on the shard chains synchronise with the Beacon chain using crosslinks. The crosslinks are the stateroots (root of Merkle tree of the shard state), and provide a means to validitate the integrity of a shard state on another shard. 

Currently, the execution in Ethereum happens in a fixed fashion using EVM, and Ethereum maintains a fixed notion of account id and balances. The plan is to generalise this notion of {\em execution environment (EE)} going forward. For instance, an execution environment using the concept of Unspent Transaction Output (UTXO) similar to Bitcoin can be hosted as one EE.  EEs could use another virtual machine in place of EVM. The EEs and users of them are distributed across multiple shards. In such a scenario, the cross-shard cross EE \eth transfers become necessary. 

This paper presents an algorithm for atomic user-level \eth transfers across EEs hosted on different shards in the proposed Ethereum upgrade. More generally, this algorithm can be used in settings where there are multiple blockchains hosting many execution environments with a mechanism to verify the integrity of the state of one chain by another chain is available. 

Because one execution environment is complete in its own right, it needs to maintain its \eth balance. An EE could provide another currency on top of \eth. Users are associated with EEs in the sense that an EE serves as a home for multiple users. The problem of transferring value from a user $a$ of EE $E_1$ to another user $b$ of another EE $E_2$ across shards involves two aspects:
\begin{enumerate}
    \item the transfer of \eth from EE $E_1$ to EE $E_2$, and
    \item the transfer of values from the user $a$ to user $b$.
\end{enumerate}

Vitalik Buterin proposed a {\em netted balance approach} (\cite{netted-balance}) for atomic cross-shard transfer of \eth between EEs. This paper proposes user-level atomic cross-shard transfer on top of this netted balance approach of EE-level transfers. A draft of this approach is published at~\cite{ethres-raghavendra}.

\section{Netted balance approach}
\label{sec:netted-balance}
Vitalik Buterin proposed a netted balance approach (~\cite{netted-balance}) for atomic cross-shard transfer of \eth between EEs. The naïve idea is for every shard to maintain its balance of every EE. In the netted-balance approach, every shard maintains the part-balances of every EE on every other shard. 

Suppose we have three shards $s_1$, $s_2$ and $s_3$. Consider an EE $E$. Then, the balance information of $E$ on shard $s_1$ is distributed on all shards $s_1$, $s_2$ and $s_3$. So, each shard now stores an ordered triple of part-balances. For example the $E$'s balance information on $s_1$ is stored as an ordered triple of the form $(10, 20, 30)$, meaning $E$'s part-balance on shard $s_1$ is 10, $E$'s part-balance on shard $s_2$ is 20, and $E$'s part-balance on shard $s_3$ is 30. Continuing the example, suppose we have triples $(-5, 10, 20)$ on $s_2$ and $(1,-2,3)$ on $s_3$. To obtain $E$'s full balance on shard $s_1$, we need to sum up $E$'s part balances on shards $s_1$, $s_2$, and $s_3$, i.e., $10 + (-5) + 1 = 6$ \eth. 

The benefit of this arrangement is that the transfer of $x$ \eth from shard $s_1$ to shard $s_2$ can be affected by an intra-shard operation only on the source shard (operation completely on $s_1$ here) by subtracting $x$ \eth from $s_1$'s part-balance and adding $x$ to $s_2$'s part-balance. The $s_1$'s triple changes from $(10, 20, 30)$ to $(10-x, 20+x, 30)$. The triples on other shards are not touched. Thus, the cross-shard EE-level \eth transfer is accomplished by a single atomic operation on a single shard.

However, the downside is that querying of an EE's balance on a shard is not a single operation because all the part balances on all other shards need to be fetched and summed. 

This idea naturally extends to cross EE transfers too.

\section{Atomic User-level Transfer}
\label{sec:atomic-user}
The core idea of this proposal is to extend and leverage the netted-balance approach of distribution of EE-balances to outstanding user-level credits and outstanding user-level reverts. The netted state (extends the netted balance) is used as a channel to communicate outstanding credits to recipient shard and outstanding reverts to the sender shard. We now describe the approach in full detail now.

\subsection{Preliminaries}

For $i$ in natural numbers, let
$s_i$'s denote shards, 
$E_i$'s denote execution environments,
$a_i, b_i$'s denote users.
We use the concept of {\em System Event messages} from~\cite{peter-cross-shard}. These messages are similar to application event messages in contract code, but are unforgeable by the application. 

In our approach, we use one System Event message called \tocredit, which includes:
\begin{itemize}
\item sender details (shard-id, EE-id, user address), 
\item recipient details (shard-id, EE-id, user address), 
\item transfer amount, 
\item the block number of the shard block where this event is emitted, and 
\item an index number starting from 0 (for every block).
\end{itemize}
We use \tocredit($a,x,b$) to denote a system event with sender details of user $a$, the transfer amount $x$, and the receipient details of the user $b$, and elide the block number and the index number when they are obvious from the context.

\subsection{Transactions}

A cross-shard transfer of $x$ \eth from an user $a_i$ on $(s_1,E_1)$ to an user $b_i$ on $(s_2,E_2)$ is split by our system into two transactions in a natural way: a cross-shard debit transfer (shown by $\Longrightarrow$) and a cross-shard credit transfer (shown by $\longrightarrow$), corresponding to deducting $x$ on the source side and adding $x$ to the destination side respectively. As is natural and expected, a user submits only a cross-shard debit transfer transaction, the corresponding cross-shard credit transfer transaction is generated by our algorithm.

{\bf Cross-shard debit transfer} transaction is signed by the sender $a_i$, and the signature is stored in the fields $v, r$ and $s$ as in   Ethereum currently. It contains a unique transaction identifier. It is submitted on sender shard, and emits a \tocredit System event on success.

{\bf Cross-shard credit transfer} transaction is submitted on recipient shard, and includes the \tocredit System Event and the Merkle Proof for it.

\subsection{Shard State}
In the netted-balance approach, each shard $s$ stores the part-balances of every EE on every shard. It can then be seen as a matrix, say $s.partBalance$ of size: number of shards $\times$ number of EE's. Here, $s.partBalance[s_i,E_j]$ gives the part-balance of EE $E_j$ on shard $s_i$, which is recorded at shard $s$. The real balance of EE $E_j$ on shard $s_i$ is given by : 
\[
	realBalance[s_i,E_j] = \sum_k s_k.partBalance[s_i,E_j].
\]

\begin{figure}[h]
	\centering
	\includegraphics[scale=0.4]{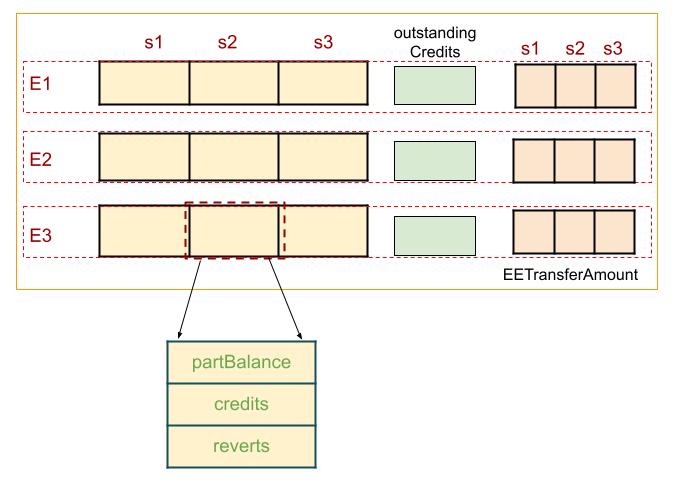}
	\caption{Shard State\label{fig:shardstate}}
\end{figure}

We introduce more elements in addition to part balances in the shard state, and we use the name $partState$ matrix in place of $partBalance$ matrix. Each cell of this matrix has 3 elements, as shown in Fig.~\ref{fig:shardstate}. They are:
\begin{itemize}
	\item $partBalance$ (as in netted-balance approach),
	\item $credits$, the set of generated cross-shard credit transactions that needs to be imported on the destination shard,
	\item $reverts$, the set of generated cross-shard revert transactions that needs to be effected on the destination shard.
\end{itemize}

The shard state contains a set called $outstandingCredits$ maintaining the cross-shard credit transactions that need to be effected on this shard. At the time of processing a block $k$ on shard $s$, for every EE $E$, $s.outstandingCredits$ is added with the generated cross-shard credit transactions targeted to $(s,E)$ from any shard $s'$ and any EE $E'$ whose corresponding debit transactions were successfully processed in the immediately previous block at shard $s'$. So, while processing block number $k$ on shard $s$, the tuple:
\[
	(s', E', k-1, \{e ~|~ e \in s'.partState[s,E].credits\}).
\]
will be added to $s.outstandingCredits$, where $k-1$ indicates that the corresponding debit transactions were processed at $k-1$ block. 

There are two ways to remove an entry from $s.outstandingCredits$. One, is by the successful proccessing of the credit transfer transaction, the other is when we {\em time-out} processing of the credit transfer. The time-out is required to guarantee a bound on the transfer time. Typically time-outs are specified in terms of number of blocks.

For every EE $E$, the shard state contains the $EETransferAmount$ that bookkeeps the net amount to be transferred at EE-level.

For every user of every EE, the shard state also stores the user's balance. We use $s.userBalance[E,a]$ to denote the balance of user $a$ of EE $E$ on shard $s$. Because the primary purpose of the shards is to divide the content, the users' balances are not distributed. Every user is assigned to a shard.

\subsection{Block Proposer Algorithm}
\label{sec:BP}
We now present the main contribution: the algorithm for the block proposer in handling cross-shard cross-EE transfers with the guarantee of atomicity. Handling of regular same-shard same-EE transfer transactions are omitted in the Step 4 of the algorithm only for the sake of succintness and clarity. This algorithm seamlessly integrates with regular transactions. 

As previously mentioned, the idea is to split a cross-shard transfer into a debit and a credit. First, a debit transaction is processed at the source shard, and the EE-level transfer is fully effected. If successful, the corresponding user-level credit transaction is queued on to the destination shard, which is processed in a subsequent block. In case of failure, the cross-shard transfer transaction needs to be resubmitted. If the credit fails, then we do the EE-level revert and user-level reverts NOT as separate transactions but as enshrined execution processing. 

Without loss of generality, assume that a Block Proposer (BP) is proposing a block numbered $k$ on shard $s_1$. Then the BP executes the steps in Fig.~\ref{fig:algo} for every EE $E_i$.

\begin{figure*}[h]
\fbox{
\begin{minipage}{\textwidth}
	\raggedright
\begin{enumerate}
\item {\bf Initialisation.} Obtain the part states of $E_i$ from every shard. Ensure that the obtained $s_i.partState$s,  $1 \le i \le n$, are correct using Merkle Proofs and crosslinks.
Compute the real balance of $E_i$ on $s_1$ using 
\[
	realBalance[s_1,E_i] = \sum_n s_n.partState[s_1,E_i].balance.
\]
For every shard $s$ and every EE $E$, set $s_1.EETransferAmount[s,E]$ to $0$.\\

\item {\bf Preprocess pending credits}
    \begin{itemize}
        \item Add entries to $s_1.outstandingCredits$
        $[s', E', (k-1)] \mapsto \bigcup_n s_n.partState[s_1,E_i].credits$
        \item Kick out expired credits from $s_1.outstandingCredits$. If there is an entry with $[s',E',k']$ such that $k' + timeOut == k$ then do:
        \begin{itemize}
            \item $s_1.partState[s',E'].reverts := s.outstandingCredits[s', E', k']$.
            \item Delete the entry with $[s', E', k']$.
			\item $s_1.EETransferAmount[s',E'] := \sum_r x_r$ where \\
			$r \in s_1.partState[s',E'].reverts$ and $x_r$ denotes $r$'s transfer amount.
        \end{itemize}
	\end{itemize}

\item {\bf Process user-level reverts}. For every $r \in s_1.partState[s_1,E_i].reverts$, update the sender's account, i.e., $s_1.userBalance[E_i,sender(r)] += x_r$, where $x_r$ denotes $r$'s transfer amount. After this operation, $s_1.partState[s_1,E_i].reverts$ is set to $\emptyset$.


\item {\bf Process transactions.} For every pair $(s_2,E_j)$
    \begin{enumerate}
	\item Select cross-shard transactions $t_1,\ldots,t_m$ between $(s_1,E_i)$ and $(s_2,E_j)$ to be included in the block. It can be a new transaction from the transaction pool, or a credit transaction from $s_1.outstandingCredits[s_2,E_j]$. 
	
	\item For every $n \in \{1, \ldots, m\}$: 
		\begin{itemize}
		\item If $t_n$ is a cross-shard debit transaction of the form $a_n \stackrel{x_n}{\Longrightarrow} b_n$ and $realBalance(s_1,E_i) > s_1.EETransferAmount(s_2,E_j) + x_n$
		\begin{itemize}
			\item include $t_n$ to the block,
			\item if $t_n$ executes successfully 
			\begin{itemize}
				\item $s_1.userBalance(E_i,a_n)$ -= $x_n$ (implied with successful execution of $t_n$)
				\item $s_1.EETransferAmount[s_2,E_j] += x_n$
				\item emit \tocredit($a_n, x_n, b_n$) System Event
				\item $s_1.partState[s_2,E_j].credits ~ \cup= ~ \{a_n \stackrel{x_n}{\longrightarrow} b_n\}$
			\end{itemize}
		\end{itemize}
		\item Else if $t_n$ is a cross-shard credit transaction of the form $b_n \stackrel{x_n}{\longrightarrow} a_n$ AND Merkle Proof check of the included \tocredit~ System Event passes AND $realBalance(s_1,E_i) > s1.EETransferAmount(s_2,E_j) + x_n$
		\begin{itemize}
			\item include $t_n$ to the block
			\item Remove $t_n$ from $s_1.outstandingCredits[s_2,E_j,k']$ where the block number $k'$ is derived from the included \tocredit ~System Event.
			\item if $t_n$ executes successfully then $s_1.usersBalance[E_i,a_n] += x_n$ (implied by the successful execution of $t_n$)
			\item if it fails \\
			$s_1.partState[s_2,E_j].reverts ~\cup=~ \{(b_n,x_n,a_n)\}$\\
			$s_1.EETransferAmount[s_2,E_j] += x_n$.
		\end{itemize}
	\end{itemize}
	\item Process EE-level transfer\\
		$s_1.partState[s_1,E_i].balance -= s1.EETransferAmount[s_2,E_j]$, \\
		$s_1.partState[s_2,E_j].balance += s1.EETransferAmount[s_2,E_j]$.
    \end{enumerate}
\end{enumerate} 
\end{minipage}}
\caption{Block Proposer Algorithm}
\label{fig:algo}
\end{figure*}

\section{Scenarios}
In this section, we apply the algorithm from Section~\ref{sec:BP} in different kinds of representative scenarios and show how atomicity is preserved.

Assume that $a_1, a_2, a_3$ are users on $(s_1,E_1)$ and $b_1,b_2,b_3$ are users on $(s_2,E_2)$.

\begin{figure*}
	\centering
	\includegraphics[scale=0.4]{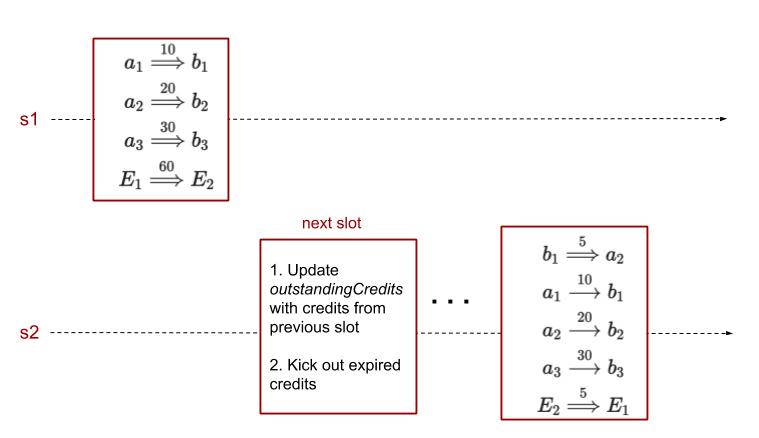}
	\caption{Happy case\label{fig:optimistic}}
\end{figure*}

\begin{figure*}
	\centering
	\includegraphics[scale=0.4]{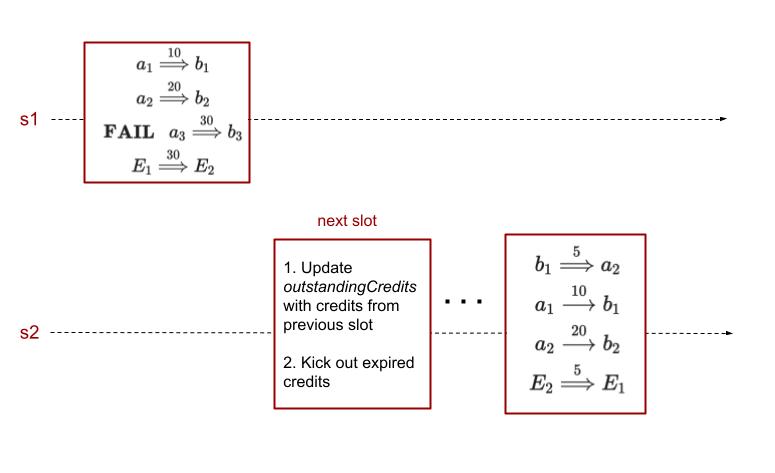}
	\caption{When debit fails\label{fig:debit}}
\end{figure*}

\begin{figure*}
	\centering
	\includegraphics[scale=0.4]{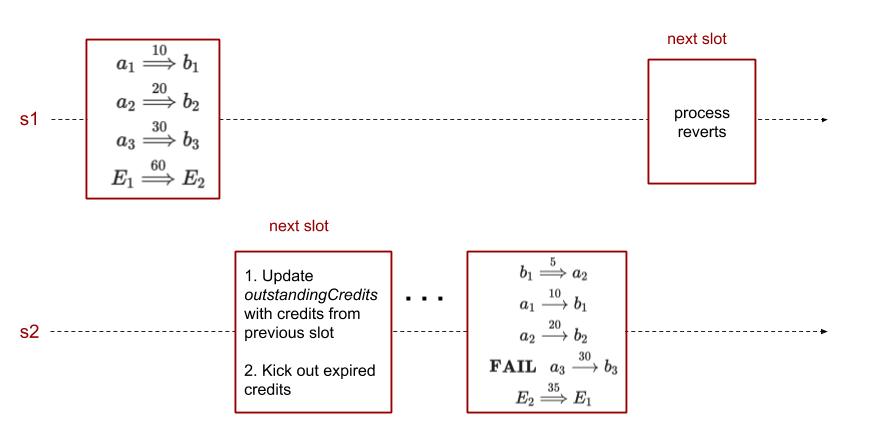}
	\caption{When credit fails\label{fig:credit}}
\end{figure*}

\subsection{Happy case}
Let us look at the happy case, where everything happens as expected. Suppose we include three cross-shard transactions as shown in Fig.~\ref{fig:optimistic} in a block on the shard $s_1$. In the very next block on $s_2$, the $s_2.outstandingCredits$ is updated with 3 pending credit transactions. In the same block or subsequently in some block on $s_2$ these credit transactions are processed. 

Note that the EE-level transfers are complete at the $s_1$'s block itself. The $s_2$'s block could process the credit transactions all in the same block or in different blocks. The blocks processing the credit transfers could include new cross-shard transactions as shown in Fig.~\ref{fig:optimistic} or other intra-shard transactions.

\subsection{When debit fails}
Consider the case when the initial debit transaction fails as shown in Fig.~\ref{fig:debit}. This is a very simple case, as nothing needs to be done. Simply this transaction has to be resubmitted as happens in Ethereum currently.

\subsection{When credit fails}
Consider a slight variation from the above scenario, where the credit transaction $a_3 \stackrel{30}{\longrightarrow} b_3$ fails or is expired as shown in Fig.~\ref{fig:credit}. Then in the same block the EE-level revert happens (with $E_1$ getting its $30$ \eth back), and finally in the very next block on $s_1$ the user-level reverts are processed.

\section{Features of the algorithm}
Some features of the algorithm are listed below.
\begin{enumerate}
\item Before processing a pending cross-shard credit transfer transaction $b_i \stackrel{x_i}{\longrightarrow} a_i$, the EE-level transfer is already complete. 
\item User-level reverts happen in the immediate next slot after a failed or an expired credit transfer. The EE-level revert happens in the same slot as the failed / expired credit transfer. This technique pushes the revert to the EE host functions instead of treating them as separate transactions. This avoids complex issues like revert timeouts and revert gas pricing.
\item Transaction identifiers need to be unique only inside the time-out number of blocks window.
\item  There is a corner case where the sender disappears by the time revert happens, then we end up in a state where there is \eth loss at user-level, but not at EE-level. We feel this is the best state to be in when such a situation happens.
\item A transaction is not included in a block if the EE does not have sufficient balance. 
\item A time-out is required to kick out long pending credit transfers. The second bullet of step 2 describes this procedure. The idea is to move out the expired user-level credit transfers, convert them into user-level reverts on the sender shard, thus achieving a fixed size for $outstandingCredits$ datastructure.
\item No locking / blocking.
\item No constraint on the block proposer to pick specific transactions or to order them.
\item {\bf Main goal} of atomicity of a cross-shard transfer is achieved.
\end{enumerate}

\section{Limitation}

In every block, the BP as well as the attestors, have to get the part-balances, outstanding credits and outstanding reverts from every other shard. This is inherited from the netted balance approach, where a BP requires the part balances from all shards. However, in the EE-level netted-balance approach the querying is restricted to only those sender EE's of the user-level transactions that are included in the block. The problem is aggravated here, because we need to query from all EE's. They (BP and attestors) either have:
\begin{enumerate}
	\item to run the full-client of all other shards, or
	\item to receive relevant parts from other shards along with their Merkle proofs over the network. 
\end{enumerate}	
Running the full clients of all other shards adds to the computation and storage overheads, while broadcasting the $partState$s of the shards after every added block adds to the network bandwidth. 

One observation is that the $partState.credits$ and $partState.reverts$ are transient, meaning, they need to store the data generated by the processing one block only. Processing of the next block overwrites this data. So, the size of $partState.credits$ and $partState.reverts$ that need to be accessed across shards is bounded by the size of the block. So larger the blocks, more is the number of inter-shard state reads, or larger are the chunks sent across the network. On Ethereum Mainnet, currently a block is produced every 13 seconds. So, one can use a storage technology supporting non-persistent faster reads to store $partState$ and alleviate this problem to a large extent.

\section{Threat Analysis of a Byzantine Block Proposer}
\label{sec:threat}
Consider the case of a Byzantine Block Proposer (BBP). A BBP might choose to deviate from the above algorithm. It becomes clear from the following that the protocol withstands such a BBP.

A validator / attester is supposed to the following checks.
\begin{enumerate}
\item Verify that the received part states from other shards for all EEs are correct.
\item Verify that the data structure $outstandingCredits$ is populated with the impending credits for this shard.
\item Verify that the impending reverts are processed, meaning the sender users are credited with the transfer amount.
\item Verify that correct \tocredit System Events are emitted for included and successful cross-shard debit transfer transactions.
\item Verify that correct outgoing credit transfers are written to the appropriate part state.
\item Verify that the outstanding credit transfer is removed when a cross-shard credit transfer happens successfully.
\item Verify that a correct revert transfer is placed for a failed cross-shard credit transfer transaction.
\item Check that the correct amount is transferred at the EE-level.
\end{enumerate}

Because a validator / an attester has access to the current shard state, (s)he can verify points: 2, 3, 5, 6, 7, 8. An attester is also given with the $partState$s from other shards along with their Merkle Proofs and (s)he has access to crosslinks form the Beacon block. So, (s)he can check points 1 and 8, that is, verify part balances, impending credits and impending reverts. Also because (s)he has access to all the transaction receipts of the transactions included in the block, (s)he can check point 4.

So, if a BBP chooses to

\begin{itemize}
	\item show no or false 
	\begin{itemize}
		\item part EE-balances or
		\item set of impending credits or
		\item set of reverts, or
	\end{itemize}
	\item not update or wrongly update $outstandingCredits$ with impending credits, or
	\item not process or wrongly process impending reverts, or
	\item not emit or emit with incorrect data the \tocredit System Event
	\item not include a revert for a failed credit transaction, or
	\item not affect appropriate EE-level transfer,
\end{itemize}
his / her block will be invalidated by the attesters, assuming that the number of Byzantine attesters are within the limit imposed by the underlying consensus algorithm.

\section{Conclusion}
We presented an atomic cross-shard cross-EE user-level value transfer algorithm for (the planned) Ethereum in presence of a Byzantine Block Proposer on top of an existing netted-balance approach for EE-level transfers. 

As part of the future work, we plan to optimise the space requirements for storing outstanding credits and outstanding reverts, and explore caching for optimising the reads of partStates of every EE of every other shard in every block (related to the above mentioned demerit).

\section*{Acknowledgements}
We thank Roberto Saltini, Peter Robinson, and Nicholas Liochon from Consensys Software R \& D, and David Hyland-Wood from Bits-Core, for all the insightful discussions.

\bibliographystyle{plain}
\bibliography{refs}
\end{document}